# Bone tumor suppression in rabbits by hyperthermia below the clinical safety limit using aligned magnetic bone cement

Xiang Yu[1†], Shan Gao[2†], Di'an Wu[1†], Zhengrui Li[1], Yan Mi[1], Tianyu Yang[1], Fan Sun[3], Lichen Wang[1], Ruoshui Liu[1], Shuli He[1,#], Qinggang Ge[4,#], Yang Lv[2,#], Andy (Yuanguang) Xu[5], Hao Zeng[3,#]

[1] Department of Physics, Capital Normal University, Beijing 100048, China
[2] Department of Orthopedics, Peking University Third Hospital, Beijing 100191, China
[3] Department of Physics, University at Buffalo, SUNY, Buffalo, NY 14260, USA
[4] Department of Intensive Care Unit, Peking University Third Hospital, Beijing 100191, China
[5] Department of Radiation Oncology, Columbia University Medical Center, New York, NY 10032, USA

## Abstract

Demonstrating highly efficient alternating current (AC) magnetic field heating of nanoparticles in physiological environments under clinically safe field parameters has remained a great challenge, hindering clinical applications of magnetic hyperthermia. In this work, we report exceptionally high loss power of magnetic bone cement under clinical safety limit of AC field parameters, incorporating DC field-aligned soft magnetic $Zn_{0.3}Fe_{2.7}O_4$ nanoparticles with low concentration. Under an AC field of 4 kA/m at 430 kHz, the aligned bone cement with 0.2 wt% nanoparticles achieved a temperature increase of 30 °C in 180 s. This amounts to a specific loss power value of 327 W/$g_{metal}$ and an intrinsic loss power of 47 nHm$^2$/kg, which is enhanced by 50-fold compared to randomly oriented samples. The high-performance magnetic bone cement allows for the demonstration of effective hyperthermia suppression of tumor growth in the bone marrow cavity of New Zealand White Rabbits subjecting to rapid cooling due to blood circulation, and significant enhancement of survival rate.

Magnetic hyperthermia is a cancer treatment technique, which utilizes the local heat generated by magnetic nanoparticles (MNPs) exposed to an AC magnetic field[1-10]. The same technique can also be used for remote neuron[11] and deep brain stimulation[12]. Specific loss power (SLP) values as high as thousands of W/g have been reported, by tuning the size[13-15], morphology[16-18] and composition[14, 19] of MNPs. For example, we obtained an SLP of 3,417 W g$^{-1}_{metal}$ at a field of 33 kA m$^{-1}$ and 380 kHz in $Co_{0.03}Mn_{0.28}Fe_{2.7}O_4/SiO_2$ MNPs[14]. In recent years, great progress has been made on *in vivo* magnetic hyperthermia treatments of solid tumors[20-27]. It was reported that tumors in mice can be eliminated after receiving magnetic hyperthermia treatment using core-shell MNPs[6]. Systemically delivered ZnMn-ferrite MNPs were shown to have increased the intratumoral temperature to above 42 °C in mice, which significantly inhibited prostate cancer growth[28]. More recently, studies also focused on heat-triggered drug release and highly efficient heat-induced immunotherapy instead of using heating alone[20, 24, 29]. In addition to injected or delivered MNPs, magnetic composite implants and magnetic scaffolds were also used for local hyperthermia[30-37]. The first magnetic hyperthermia was attempted in 1957 for metastasis in lymph nodes[1]. In 2005, the first clinical application of interstitial hyperthermia using MNPs in locally recurrent prostate cancer was reported[38]. Maier-Hauff *et al.* published results from a Phase I clinical study involving 14 patients with recurrent glioblastoma multiforme[39, 40]. These clinical studies were performed using a commercial machine (MagForce MFH 300F). A roadmap for magnetic hyperthermia has been published earlier this year, providing a review of the current state-of-the-art while also pointing out significant challenges[41].

Clinical magnetic hyperthermia poses stringent requirements for the heating performance of MNP materials.[42, 43] First, in a clinical environment, the motion of MNPs is often impeded or frozen; second, there exists fast heat dissipation due to the blood circulation; and third, clinical safety concerns put a limit on the AC field parameters and concentration of MNPs that can be used. The high SLP values reported in earlier studies were typically obtained in fluids with freely rotating MNPs, often under quasi-adiabatic conditions, and with AC field parameters unsuitable for clinical applications, and therefore cannot directly translate to heating performance in clinical settings. In an earlier clinical study, the tolerable field amplitude was found to be 6 kA/m for the pelvic region, 7.5 kA/m for thoracic and neck region, and up to 13.5 kA/m for the head/brain region, at a frequency of 100 kHz[40]. Later, the clinical safety limit for the product of AC field amplitude and frequency was proposed to be $Hf < 5 \times 10^9$ A/(m·s) to avoid direct tissue heating [3, 44]. A lower $Hf$ product also makes full body magnetic hyperthermia economical. In the past few years, a number of papers reported the heating performance of MNPs embedded in bone cement and epoxy to mimic frozen MNPs in vivo[45-49]. The reported SLP values ranged from 10 W/g at 4 KA/m, 100 kHz (intrinsic loss power, ILP= 5.7 nHm$^2$/kg) to 122 W/g at 24 kA/m, 536.5 kHz (ILP= 0.3 nHm$^2$/kg). One group reported erroneously high SLP of hundreds of W/g; based on the heating curves provided, the actual SLP is estimated to be less than 60 W/g at 5.72 kA/m and 626 kHz [50, 51]. These studies suggest that under realistic clinical conditions (immobilized MNPs, fast heat dissipation and $Hf < 5 \times 10^9$ A/(m·s)), high performance MNP materials suitable for hyperthermia remain elusive.

In this work, magnetic bone cement with unprecedented *in vivo* heating performance under clinically safe AC field parameters was developed, by embedding low concentration of monodisperse, 22 nm soft-magnetic $Zn_{0.3}Fe_{2.7}O_4$ (ZFO) MNPs in PMMA bone cement. Application of a DC field during curing of magnetic bone cement results in long chain formation of MNPs and reorientation of their magnetic easy axes. The magnetic bone cement containing only 0.2 wt% MNPs exhibits *in vitro* heating by 30 °C in 180 s, achieving an SLP value of 327 W/g$_{metal}$ under a low AC field of 4 kA/m at 430 kHz ($Hf = 1.7 \times 10^9 \, A/(m \cdot s)$), and an ILP of 47 nHm$^2$/kg. These values represent 50-fold enhancement compared to random magnetic bone cement. The random magnetic bone cement produced insignificant heating under identical conditions, despite the fact that the same MNPs exhibited highly efficient low field heating in aqueous solutions[14]. *In vivo* experiments showed that the bone cement containing 2 wt% MNPs implanted in the bone marrow cavity of the tibia of New Zealand White Rabbits can be heated to a steady temperature of 54 °C, under an AC field of 7 kA/m and 380 kHz ($Hf < 2.7 \times 10^9$ A/(m·s)), despite fast heat dissipation from blood circulation. Effective tumor growth suppression and significant enhancement of survival rate were demonstrated, using VX2 tumor model inoculated in the bone marrow cavity of the tibia of rabbits. Such magnetic bone cement can be used for cancer treatment to supplement or even replace chemo/radiation therapy after surgical removal of tumor. Due to the low concentration and bio-compatibility of MNPs embedded in bone cement, elimination of MNPs post treatment may no longer be necessary. Our work thus represents a transformative approach towards clinical magnetic hyperthermia applications.

Fig. 1(a) and 1(b) show typical transmission electron microscopy (TEM) images of ZFO MNPs with diameters of 16 nm and 22 nm, respectively. It can be seen that these MNPs are monodisperse with narrow size distribution (shown in Fig. S1). A high-resolution TEM image (Fig. 1 (c)) shows that the MNP is single-crystalline, with (220) lattice fringes showing a spacing of 2.9 Å, consistent with those of iron ferrites. A selected area electron diffraction pattern shown in Fig. 1(d) confirms the MNPs to possess a spinel structure. The XRD patterns of the MNPs with sizes of 16, 18, 20 and 22 nm are shown in Fig. S2, in agreement with their corresponding standard JCPD card (PDF#86-0510)[52]. The average sizes of MNPs were estimated by Scherrer's formula. The estimated sizes are consistent with TEM images. The specific composition and sizes were chosen because such a combination was optimized for low field hyperthermia applications, showing a high SLP of 500 W/$g_{metal}$ and an ILP of 26.8 nHm$^2$/kg in aqueous solution, at an AC field of 7 kA/m and 380 kHz, in our previous work[14].

In clinical hyperthermia, the physical motion of MNPs is restricted due to their anchoring on cell surfaces, which could result in vastly different AC field heating behaviors[53-55]. To investigate the effect of immobilization, 0.2 wt% MNPs were dispersed in PMMA to form a magnetic bone cement, with or without applying a 2 T DC field during immobilization. Fig. 1(e) and (f) are TEM images of ZFO MNPs in PMMA without and with applying the DC field, respectively (more images are shown in Fig. S3). As can be seen from Fig. 1(e), without field alignment, the MNPs form a random assembly, with short chains of MNPs orienting in random directions. With a DC field applied (Fig. 1(f)), however, the MNPs form long and parallel chain bundles to minimize the magnetostatic energy.

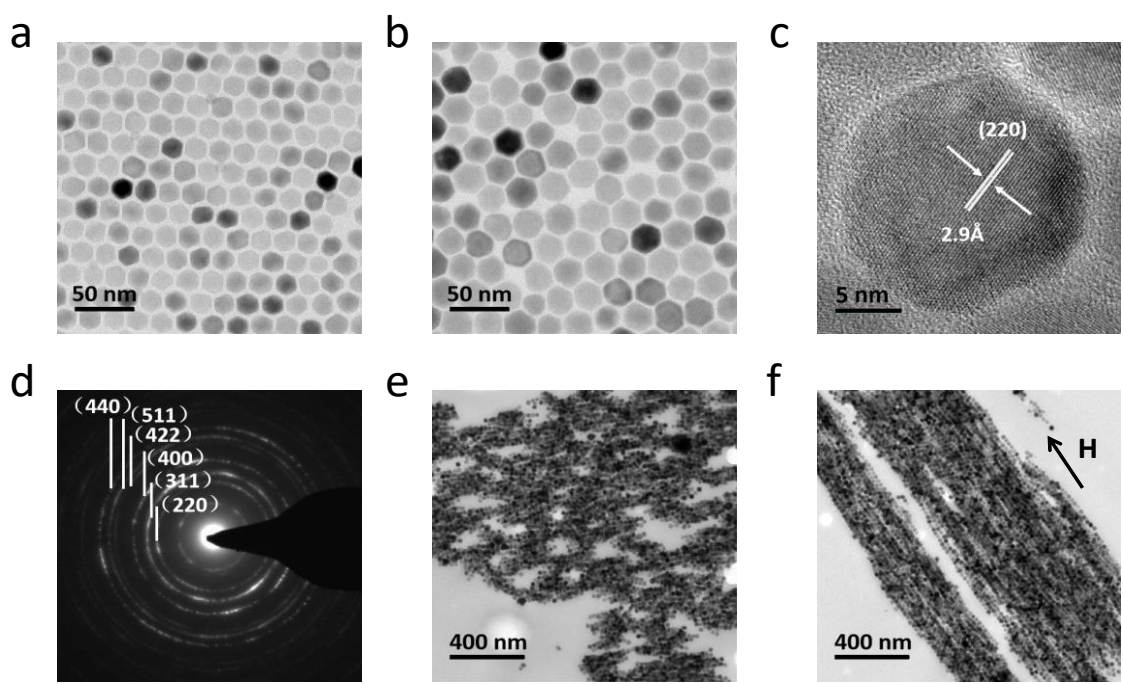

**Figure 1.** The TEM images of ZFO MNPs with sizes of (a)16 nm and (b) 22 nm; (c) a HRTEM image of a 22 nm-ZFO MNP; (d) SAED pattern acquired from a 22 nm-ZFO MNP assembly; the TEM images of (e) a random and (f) an aligned magnetic bone cement containing 0.2 wt% 22 nm-ZFO MNPs.

The DC magnetic hysteresis loops (shown in Fig. S4) reveal that the ZFO MNPs of 16-22 nm are all superparamagnetic at room temperature. The surfactant remaining on the MNPs was determined to be 10 wt% by thermogravimetric analysis (TGA), which was subtracted to obtain the net magnetization. As the diameter of MNPs increases from 16 nm to 22 nm, saturation magnetization ($M_S$) increases from 440 kA/m to 461 kA/m (Fig. S4). This moderate increase is due to the reduced volume fraction of canted surface spins with increasing size[56]. To understand the AC field heating behavior, the minor hysteresis loops with field amplitudes of 4, 10, 14 and 27 kA/m, in the same range to those used for hyperthermia experiments, were measured for magnetic bone cement with random and aligned 22 nm MNPs, respectively. Fig. 2 shows the minor hysteresis loops measured at 20, 100, 200 and 250 K, with (top panel) and without (bottom panel) field alignment, respectively (the measuring field is applied in the direction of alignment). It can be seen that the random and aligned magnetic bone cement exhibit substantially different loop shapes. The aligned samples show squarer loop shape with much higher remanence ratio ($M_r/M_s$), where $M_r$ is the remanent magnetization. In contrast, the random samples exhibit slanted hysteresis loops with tiny opening[57, 58]. On the other hand, the aligned samples exhibit higher coercivity at low temperatures than that of random ones, which then becomes similar at higher temperatures above 100 K. As a result, the areas enclosed by the hysteresis loops, which determine the SLP values, are much larger for the aligned samples than for the random ones. This is due to the field alignment during the curing process of the bone cement. When an external field is applied during drying, not only the MNPs form long chains to minimize the magnetostatic energy, but they also physically rotate to align their easy axes in the field direction (*i.e.* along the chain axis) to minimize the Zeeman energy. These two effects work synergistically to increase the area of the hysteresis loop [59]: the alignment of the easy axes sharply increases the remanence ratio; while the chain formation leads to steeper transition near the coercive field. In a one dimensional chain of MNPs, the magnetization reversal proceeds by flipping the moment of a particle, followed by propagation along the chain aided by the dipole field from the flipped MNP. Thus the switching field distribution is substantially narrower compared to random samples, as seen in Fig. 2. Both the enhanced ramanence ratio and narrower switching field distribution should translate to higher SLP values for the aligned samples[60]. In contrast, for the random samples, at low field amplitudes (4-7 kA/m) that are smaller than the anisotropy field, misalignment of the easy axes of MNPs from the AC field direction leads to inability to flip their magnetic moment. The magnetization is then nearly linear in field, as shown in the bottom panel of Fig. 2, which would lead to greatly diminished SLP.

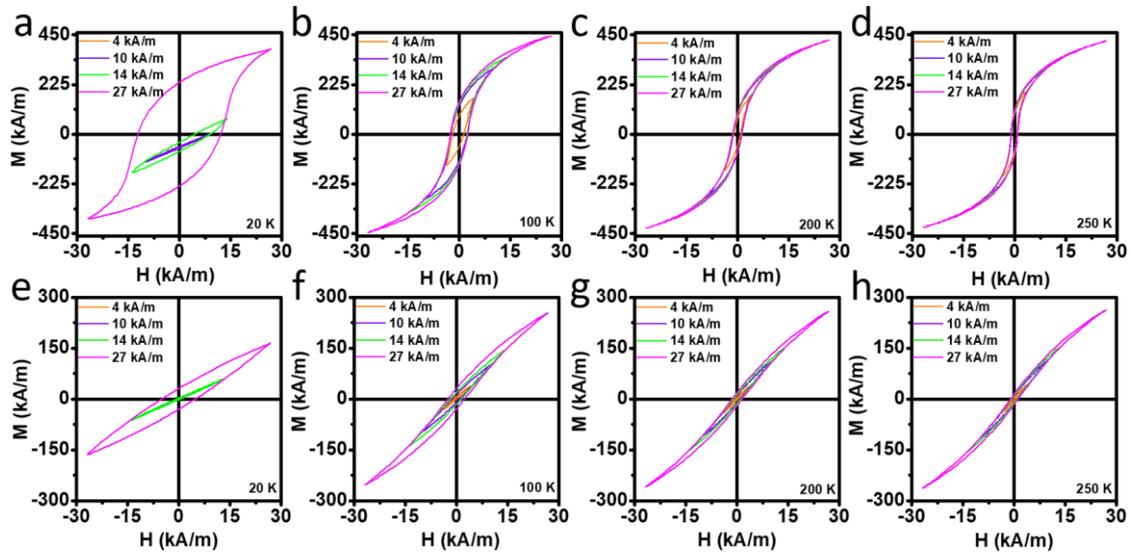

**Figure 2.** Minor hysteresis loops of aligned magnetic bone cement containing 0.2 wt% 22 nm ZFO MNPs measured at temperatures of (a) 20 K, (b) 100 K, (c) 200 K and (d) 250 K; minor hysteresis loops of random magnetic bone cement measured at temperatures of (e) 20 K, (f) 100 K, (g) 200 K, and (h) 250 K. The vertical axis represents the magnetization of the ZFO MNPs.

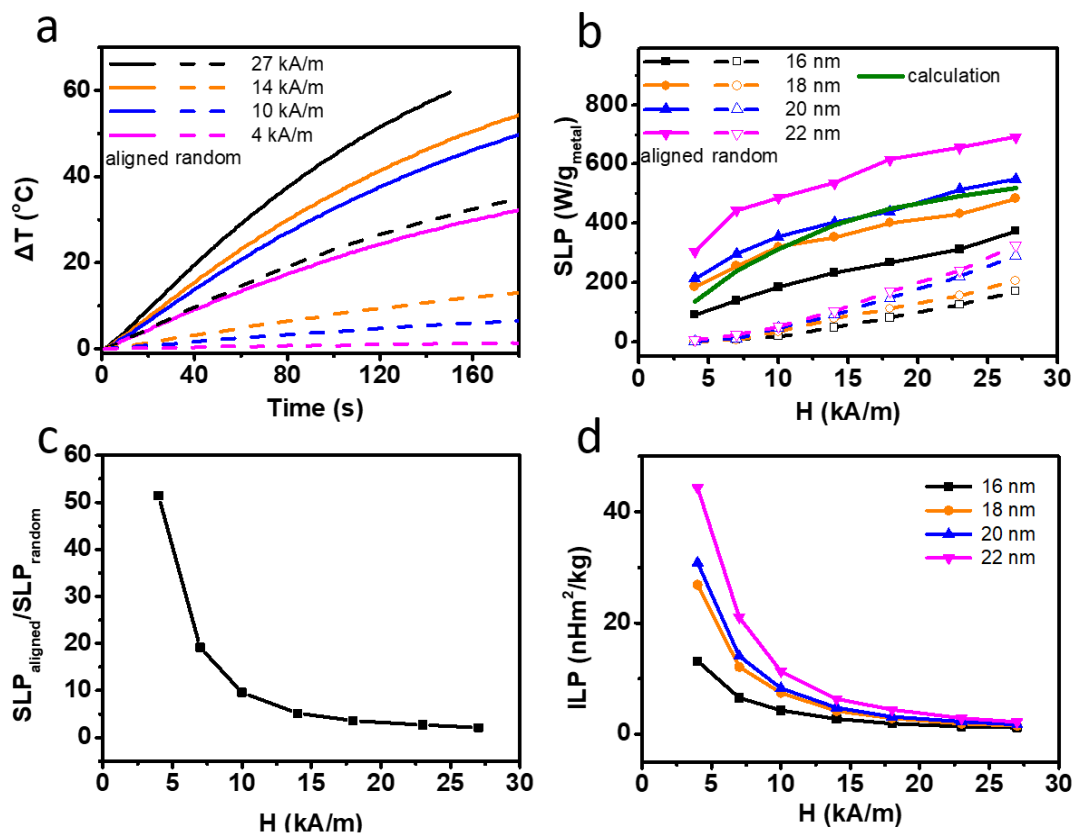

**Figure 3.** (a) Representative AC field heating curves of aligned (solid lines) and random (dashed lines) magnetic bone cement containing 0.2 wt% 22 nm-ZFO MNPs, measured at 4, 10, 14 and 27 kA/m, respectively; (b) Measured SLP of aligned (solid lines) and random (dashed lines) samples as a function of field amplitude, respectively. Also plotted is the SLP estimated from the minor hysteresis loops at 100 K using Eq. (1) (green line); (c) The ratio of SLP of aligned over random samples as a function of field amplitude; and (d) ILP as a function of field amplitude of aligned samples containing 0.2 wt% ZFO MNPs with diameters of 16, 18, 20, 22 nm, respectively.

Fig. 3(a) shows the AC field heating curves of aligned (solid lines) and random (dashed lines) magnetic bone cement incorporating 0.2 wt% 22 nm-ZFO MNPs under the field frequency of 430 kHz and amplitudes ranging from 4, 10, 14 and 27 kA/m, respectively. The heating curves at all measured field amplitudes from 4-27 kA/m are shown in Fig. S5. Indeed, the magnetic bone cement with random MNP orientation showed negligible heating at 4 and 7 kA/m, as seen from Fig. 3(a), despite these MNPs being optimized for low field heating in aqueous solutions, and our early work showing that they exhibit high SLP of 500 W/g at 7 kA/m[14]. Strikingly, for the aligned 22 nm sample, the temperature increases rapidly by 30 °C in 180 s, even under an ultralow field of 4 kA/m ($Hf = 1.72 \times 10^9$ A/(m·s)). For comparison, similar temperature increase is only achieved for the random sample at a significantly larger field of 27 kA/m. As a comparison, a recent study showed that it took 6% (30-fold higher concentration) randomly oriented $Fe_3O_4$ MNPs in PMMA under a higher $Hf$ value of $3.58 \times 10^9$ A/(m·s) (5.72 kA/m and 626 kHz) to reach similar temperature increase[51]. This shows that our aligned magnetic bone cement exhibits far superior AC field heating performance, suggesting that field alignment is highly effective in boosting the heating capacity, especially at ultralow field amplitudes.

SLP as a function of field amplitude for the aligned (solid lines) and random (dashed lines) bone cement with MNP sizes ranging from 16 to 22 nm are plotted in Fig. 3(b). It can be seen that at 4 kA/m, while SLP is only 6 W/g for the random sample with 22 nm MNPs, it reaches 327 W/g for the aligned sample with 22 nm MNPs. Heating curves of random and aligned samples with MNP sizes ranging from 16 to 20 nm are shown in Fig. S6 (a)-(f). It can be seen that all the random samples exhibit negligible heating under the field amplitude of 4 kA/m, while the temperature of aligned samples can be increased by 15-30 °C in 180 s.

The origin of AC field heating is hysteresis loss. Although MNPs are superparamagnetic in quasi-static measurements, they have to possess a dynamic hysteresis loop with a non-zero area to produce heat loss. This can be understood in the simplified framework of Stoner-Wohlfarth model, which describes the hysteresis behavior of non-interacting MNPs with uniaxial anisotropy undergoing coherent rotation. According to this model, the coercivity $H_c$ of the MNP ensemble can be described by

$$H_C(T) = H_{C0}(T)\left[1 - \left(\frac{k_B T ln(f_0\tau)}{K_u(T)V}\right)^\alpha\right] \quad (1),$$

where $H_{C0}(T)$ is the DC coercivity at absolute temperature T, $k_B$ is the Boltzmann constant, $f_0$ is the attempt frequency (typically taken to be $10^9$ Hz), $\tau$ is the measurement time, $K_u$ is the anisotropy constant and V is the magnetic volume of a single MNP. $\alpha$ is taken to be 2/3 or 3/4 for randomly oriented ensemble and 1/2 for aligned ensemble [61-64]. According to Eq. (1), MNPs can be superparamagnetic under DC conditions but possess a frequency-dependent dynamic $H_C$. However, dynamic hysteresis loops can only be measured in limited field amplitude and frequency ranges by specialized instrumentation[16, 19, 65]. As an alternative, we adopt the following approach to estimate the dynamic hysteresis behavior and SLP of the magnetic bone cement (at 300 K): First, we find the temperature $T_2$ (lower than 300 K) at which the quasi-static $H_{C2}$ (measured by VSM with $f_2 \approx 0.01\ Hz$) is equal to the dynamic $H_{C1}$ ($f_1 = 430\ kHz$) at $T_1 = 300$ K using Eq. (1) (see details in Supporting Information). Second, we use the minor hysteresis loop area measured at $T_2$ to represent the dynamic hysteresis loop area at $T_1$, and the loss power is then calculated as $SLP = S \cdot f$, where $S$ is the area of the minor hysteresis loops measured at $T_2$ with field amplitudes from 4-27 kA/m. Considering also the temperature dependence of $H_{C0}$ and $K_u$[66], $T_2$ is estimated to be ~ 100 K. The SLP as a function of field amplitude for the aligned bone cement with 22 nm MNPs, estimated from the minor hysteresis loops measured at 100 K (Fig. 2(c)), is plotted in Fig. 3(b) (the green curve). It can be seen that the estimated SLP vs field qualitatively matches the trend of the measured curve, validating our approach. Quantitative comparison was not attempted due to the simplification of the model ignoring interactions and the uncertainties in temperature dependence of material parameters.

The enhancement factor, defined as the ratio of SLP of aligned and random samples, was calculated to quantitatively compare the performance of aligned and random samples. It is plotted as a function of field amplitude in Fig. 3(c). The enhancement factor reaches 50 at 4 kA/m, which then decreases monotonically with increasing field. At the field of 27 kA/m, there is still an impressive enhancement by > 100%. It is exciting that the SLP of the oriented sample is nearly 50 folds of magnitude higher than that of the random sample at the low fields, as this is critical for enabling clinical hyperthermia which necessitates low fields and low MNP concentrations. The ILP of oriented cement as a function of field amplitude for MNP sizes of 16-22 nm are plotted in Fig. 3(d). It can be seen that the ILP for the sample with 22 nm MNPs is 47 nHm$^2$/kg at 4 kA/m, such a value is unprecedented for MNPs absence of rotational degree of freedom.

It is expected that MNPs fixed in a medium will show lower SLP than aqueous solutions, due to the freezing of Brownian motion. However, it is still surprising to observe SLP decreasing by orders of magnitude, *e.g.* from 500 W/g to 20 W/g at 7 kA/m[14], especially considering that these MNPs were optimized for low field applications. From our previous experiments on SLP as a function of viscosity using water: glycerin mixtures, we estimate that Brownian motion contributes to ~ 50% of the SLP for this type of MNPs. Thus locking out Brownian motion alone cannot explain the difference. As discussed above, the main reason for the low SLP of the random sample is the misalignment of the easy axes of MNPs leading to inability to flip their magnetic moment at low fields. On the other hand, in an aqueous or viscos medium, the MNPs can reorient their easy axes in the direction of the AC field after field cycling, thus increase their magnetic softness overtime. In earlier studies, a delay of the onset of heating by a few seconds was often observed, but not clearly explained[14, 65]. We suggest that this delay is due to the time it takes for the MNPs to reorient their easy axes. Since the physical rotation of MNPs is typically restricted in clinical settings, SLP measured in an aqueous solution can overestimate the actual heating performance by orders of magnitude. Treatment planning such as choices of dosage and field parameters thus should never be based on SLP measured in aqueous solutions. Take the most investigated material for hyperthermia, magnetite ($Fe_3O_4$) for example, its $K_u$ is ~ $8.5\times10^3$ J/m$^3$ and its anisotropy field is ~ 20 kA/m. This explains why a high concentration of MNPs was required to obtain sufficient temperature rise in clinical hyperthermia with fields smaller than 20 kA/m[40]. We further suggest that attempts to enhancing SLP by increasing the anisotropy of $Fe_3O_4$ MNPs through composition, shape and core/shell tuning will be futile for clinical applications requiring low fields. Research should instead focus on bio-compatible, soft magnetic MNPs such as zinc ferrites used in the present work.

Fig. 4(a) and 4(b) show the IR images of the pig rib filled with aligned and random magnetic bone cement subjected to an AC field of 4 kA/m at 430 kHz. It can be seen that the temperature of the oriented magnetic bone cement rises rapidly to above 50 °C within 60 s, and further to 85 °C at 600 s. Meanwhile, the temperature of the bone 7 mm away from the center reaches above 45 °C at 600 s. However, there is no measurable heating of the random magnetic bone cement.

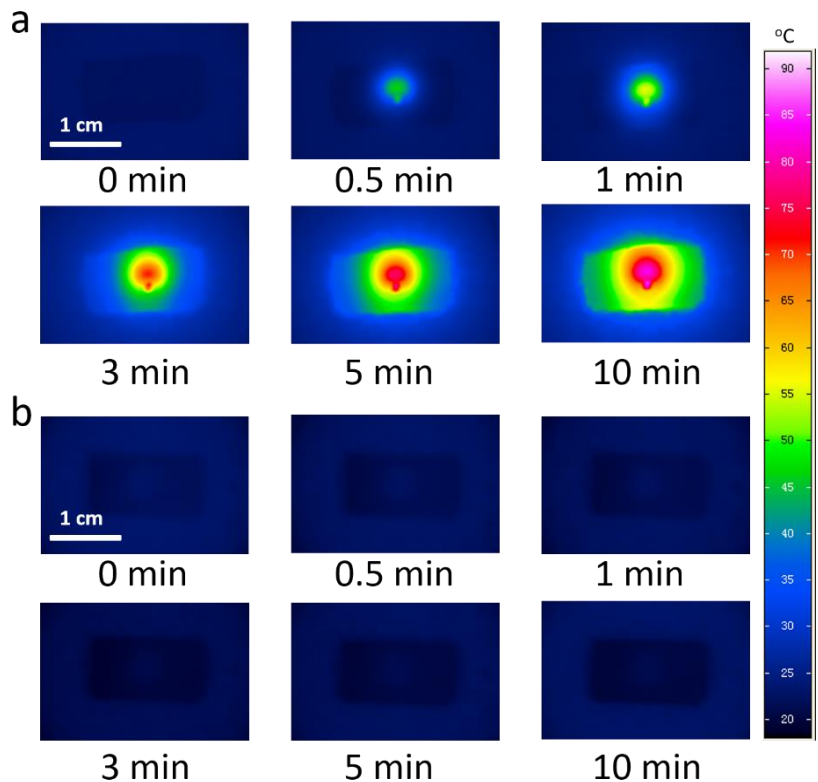

**Figure 4.** Infrared images of the (a) aligned and (b) random magnetic bone cement subjected to an AC field of 4 kA/m at 430 kHz.

Cytotoxicity of bone cement containing 1% and 2% of ZFO MNPs was also measured and compared to PMMA bone cement without MNP and blank control. The mouse Embryonic fibroblast (MEF) cell lines were chosen for *in vitro* experiments according to the ISO 10993-14. As shown in Fig. 5(a), the cell viability after incubation for 24, 48 and 72 hr., of both 1% and 2% ZFO bone cement groups, shows no significant difference (P > 0.05) from the blank control and PMMA bone cement groups, demonstrating the excellent biocompatibility of the magnetic bone cement ZFO MNPs. This is also consistent with our prior work showing biocompatibility of ZFO MNPs using MEF and MG-63 cell lines [13].

After demonstrating exceptional heating performance of the aligned magnetic bone cement *in vitro*, the suitability of such material for clinical hyperthermia applications was investigated. *In vivo* AC field heating measurements were carried out in the tibial medullary canal of New Zealand White Rabbits. A magnetic bone cement rod was surgically implanted into the tibial medullary canal and subjected to an AC field of 4 kA/m to 25 kA/m at 380 kHz, and the temperature of the medullary canal was monitored by a fiberoptic thermometer (Fig. 5(b)). As shown in Fig. 5(c), the medullary canal cannot be heated by the random magnetic bone cement at field amplitudes of 4 kA/m and 7 kA/m. The temperature drop from 32 to 29 °C is not an artifact but due to the decline of body temperature after anesthesia. For the aligned magnetic bone cement, temperature increases of 15 °C and 21 °C were found at 4 kA/m and 7 kA/m, respectively, at 1500 seconds (Fig. 5(d)). While these temperatures are suitable for hyperthermia applications, the temperature increase is significantly slower and the maximum temperature reached is lower than *in vitro* results. This is primarily because the body temperature of the rabbit is regulated by blood flow to keep it nearly constant, which leads to significantly greater heat dissipation compared to *in vitro* environment. We note that the smaller dimension of the bone cement used for *in vivo* experiments and the slightly lower frequency (due to the larger coil needed for *in vivo* measurements) also lower the heating rate and achievable temperature. We suggest that using self-limiting MNPs with low Curie temperature for hyperthermia is misleading as they would have poor performance due to the low field and fast heat dissipation in clinical settings[67]. With increasing field amplitude to 25 kA/m, it takes only 150 seconds to reach a temperature rise of 21°C. Through this *in vivo* experiment, it is confirmed that the aligned magnetic bone cement can reach the temperature needed to carry out mild magnetic hyperthermia at an ultralow field of 4 kA/m; and the temperature needed for thermal ablation ($\Delta T > 20$ °C) can be reached under the clinical safety limit ($H = 7$ *kA/m, f* $=380$ *kHz, Hf* $= 2.67 \times 10^9$ A/(m·s)).

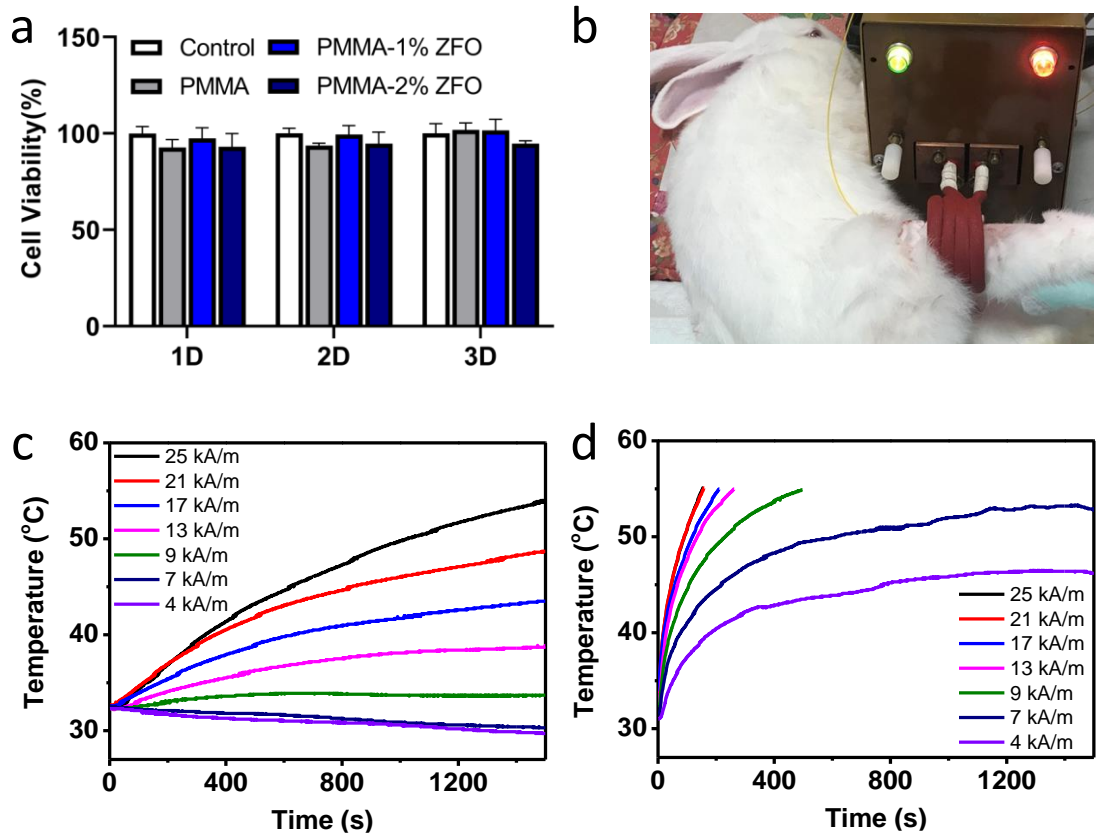

**Figure 5.** (a) Cytotoxicity of bone cement containing 1 wt. % and 2 wt.% ZFO MNPs to MEF cells; (b) A picture of the New Zealand White Rabbit used for the *in vivo* AC field heating experiments. Heating curves of (c) random and (d) aligned magnetic bone cement containing 2 wt.% ZFO MNPs in the medullary canal of the New Zealand White Rabbit.

We next established a bone tumor model to simulate bone tumor invasion and verify the efficacy of magnetic hyperthermia (Fig. S8). First, we investigated the growth of VX2 tumor in rabbit tibia by magnetic resonance imaging (MRI). MRI results showed that VX2 tumor masses became active when the tumor had been seeded in rabbit tibial on postoperative day 3, indicating VX2 tumor models were successfully established (Fig. S9). Therefore, magnetic hyperthermia treatment (7 kA/m at 380 kHz) of rabbit tibial was performed starting on postoperative day 3. The treatment lasted 30 min each for 4 days. Fig. 6(a) shows the reconstructed CT images at each follow-up time point. The tumor regions are marked by red circles. The magnetic bone cement rod used for heating of the bone marrow cavity, with one end located near the mid tibia, and a second short rod used for sealing the hole, are clearly visible. Based on CT results on postoperative day 3, the bone structures of both groups remain intact. After 4 days of hyperthermia treatment (postoperative day 7), the cortical bone appears complete with no obvious damage to the normal bone tissue. CT results in the control group suggests that the tumor tissue had not yet invaded the normal bone tissue. On the 14th postoperative day, the control group showed significant cortical bone destruction (red arrow). The cortical bone near the tibial tuberosity started to show bone density loss and the proximal periosteum was pushed outward forming a tumor-induced bone structure. The time of emergence of bone destruction was found to be 2 weeks. While for the treatment group, no obvious abnormality was found compared with the previous testing time. The cortical bone destruction was more pronounced in the control group at 28 days postoperative. There was apparently spongy structure formation in the cortical bone from the beginning of the tibial tuberosity to the middle tibia and indication of cancer invasion to the surrounding tissues. The tumor tissue infiltrated the space between the heating rod and sealing rod, and invaded the contralateral bone tissue. Whereas there was no significant abnormality observed in the treatment group at 28 days postoperative. The results comparing the 10 rabbits each in the treatment and control groups are qualitatively similar to those shown in Fig. 6 (a). These CT images are available upon request.

Survival study was performed on both treatment and control groups. Fig. 6 (b) plots the Kaplan−Meier survival curves showing the survival rate of rabbits following hyperthermia treatment and without treatment. According to ethical regulations, the observation period of the study was set as 84 days. During the observation period, all animals in the control group died, and the median survival time was 30.5 days (shown by the blue dashed line). In the treatment group, 5 animals (50%) survived at the end of the observation period, and the median survival time was 80.5 days (red curve). The difference was statistically significant (P < 0.0001). Compared to the control groups, the treatment group exhibited a significant survival advantage, with 100% survival for at least 60 days after hyperthermia treatment. It is clear that the hyperthermia therapy improved the survival rate of rabbits substantially.

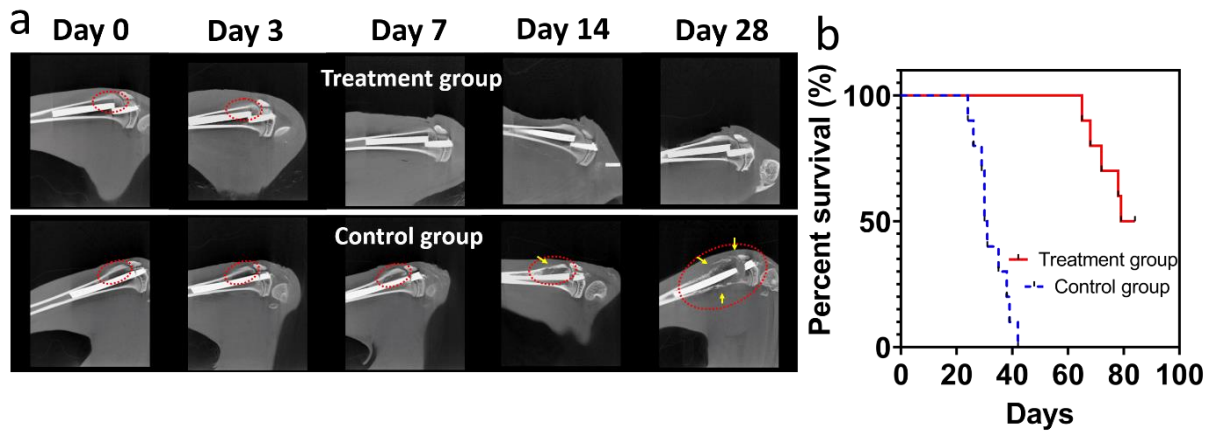

Figure 6 (a) Reconstructed CT images of rabbit legs at each follow-up time point for treatment group (top panel) and control group (bottom panel). Red circles represent the regions with tumor; yellow arrows indicate cortical bond destruction; (b) Survival rate of rabbits bearing tumors after magnetic hyperthermia treatment and in the control group.

In this work, magnetically soft $Zn_{0.3}Fe_{2.7}O_4$ nanoparticles with 16 nm, 18 nm, 20 nm and 22 nm were prepared by thermal decomposition method. These MNPs were incorporated into PMMA bone cement and aligned in a DC magnetic field to address the stringent requirements for clinical hyperthermia, namely low AC field amplitude, fast heat dissipation, low MNP concentration and frozen MNP motion. Our approach yields unprecedented SLP and ILP values in aligned samples, which exhibit 50-fold enhancement compared to random samples at a low field amplitude of 4 kA/m. This is explained by the synergistic effect of chaining of MNPs and alignment of their easy axes along the chain axis, which substantially increases the area of the AC hysteresis that dictates the power loss. Our approach enabled the magnetic bone cement to demonstrate high efficacy for hyperthermia cancer treatment *in vivo*. Using VX2 tumor model in the tibial medullary canal of New Zealand White Rabbits, our material demonstrated adequate *in vivo* temperature increase and subsequent tumor suppression, leading to significantly enhanced survival rate after the hyperthermia treatment.

Magnetic hyperthermia is a local and controllable cancer treatment. With lower side effects than chemo/radiation therapy and no accumulated toxicity, patients can tolerate frequent treatments. The treatment can kill drug resistant tumor cells, which is the most challenging for chemotherapy. Due to the low concentration of MNPs used in the magnetic bone cement, the presence of MNPs should not affect the biocompatibility, mechanical properties, and operability. The MNPs embedded in the bone cement is not expected to migrate to other parts of the body, and may no longer need to be eliminated from body post treatment. Our work calls for future clinical studies to determine the optimal treatment dosage (temperature-time) for different tumor types, as our high performance material offers possibilities for both mild hyperthermia and thermal ablation.

*Materials*

Fe(III) acetylacetonate (99%) was purchased from Alfa aesar; Zn(II) acetylacetonate (99%), oleic acid (90%) and oleylamine (70%) was purchased from Sigma-Aldrich; dodecanediol (93%) was purchased from TCI; benzyl ether (99%) was purchased from Acros; ethanol (99%) and hexane (97%) were purchased from Beijing Tong Guang Fine Chemicals Company.

*Synthesis of $Zn_{0.3}Fe_{2.7}O_4$ nanoparticles*

ZFO MNPs with a diameter of 6 nm was synthesized by thermal decomposition in organic solutions[68, 69]. Briefly, Fe(III) acetylacetonate (2.7 mmol, 0.9536 g), Zn(II) acetylacetonate (0.3 mmol, 0.079 g), dodecanediol (10 mmol, 2.0234 g ), oleic acid (3 mmol, 0.96 mL), oleylamine (3 mmol, 1.02 mL), and 20 mL benzyl ether were mixed and magnetically stirred (500 rpm) under a flow of nitrogen (30 mL/min). The mixture was heated to 473K for 1 h and further heated to reflux (~ 573 K) for 2 h. The solution was cooled down to room temperature after removing the heat source. The MNPs were collected by adding ethanol and centrifugation (6500 rpm, 5 min), and were stored in hexane. ZFO MNPs with a diameter of 22 nm were synthesized following a seed-mediated growth technique reported by Sun and Zeng[69]. Details are provided in the Supporting Information.

*Preparation of random and aligned magnetic bone cement*

A bone cement was made from polymethyl methacrylate (PMMA) powder and methyl methacrylate (MMA) liquid. To prepare magnetic bone cement, 3.7 mg ZFO MNPs were fully dispersed in 0.75 ml MMA liquid by ultrasonication. The mixture was fully mixed with 1 g PMMA powder by stirring. The bone cement fluid was injected into a cylindrical alumina crucible. A DC magnetic field of 2 T was applied along the axis of the alumina mold. After solidification for 40 minutes, aligned magnetic bone cement was obtained. Without applying a magnetic field, random magnetic bone cement was prepared. The magnetic bone cement used for *in vitro* experiments is of the shape of a cylindrical rod with a diameter of 6 mm and length of 10 mm and contains 0.2 wt% of MNPs; the magnetic bone cement used for *in vivo* experiments is 3 mm in diameter and 35-40 mm in length, containing 2 wt% of MNPs. Due to the smaller diameter of the bone cement required to be used in the medullary canal of the rabbit tibia and fast heat dissipation *in vivo*, a higher concentration of MNPs is necessary.

*Characterizations*

The size, morphology and crystal structure of the MNPs were studied by a high-resolution transmission electron microscope (FEI Tecnai G2 F30) and selected area electron diffraction. Energy dispersive X-ray spectroscopy element mapping was performed by a scanning electron microscope (Hitachi S-4800). Crystal structure was further analyzed using an X-ray diffractometer (Bruker D8 Advance). TGA was performed on MNPs to subtract the organic coating content (Netzsch TG 209 F3 Tarsus). Magnetic hysteresis loops were measured by a vibrating sample magnetometer (GMW Model3473-70) and a superconducting quantum interference device magnetometer (Quantum Design SQUID-VSM). Hyperthermia performance of the magnetic bone cement containing aligned and random MNPs was investigated by a HYPER5 machine (MSI Company) under an AC magnetic field with amplitude ranging from 4-27 kA/m and frequencies of 430 kHz (for *in vitro* measurements) and 380 kHz (for *in vivo* treatment due to the larger sized coil to accommodate rabbit tibia), respectively. The temperature change of the MNP solution and magnetic bone cement was monitored by a fiberoptic probe. To test the ability of the magnetic bone cement to heat the surrounding bone structure, a piece of pig rib with a hole of 6 mm in diameter and 10 mm in length was filled with oriented and random magnetic bone cement, and subjected to an AC field of 4 kA/m at 430 kHz. The temperature rise of the pig rib was recorded by a high-resolution infrared (IR) camera.

*Cell culture*

Fibroblast L929 cells were purchased from iCell Bioscience Incorporated (Shanghai, China). Cells were cultured in MEM (Gibco, Invitrogen, USA) supplemented with 10% (v/v) equine serum (Hyclone, USA) and antibiotics (10 μg/mL penicillin and 10 μg/mL streptomycin). Cells were incubated at 37 °C in a humidified atmosphere of 5% $CO_2$ until approximately 80% confluence was reached. They were then trypsinized in 0.25% pancreatic enzymes before being suspended in fresh culture medium. Cells were counted using a haemocytometer and an inverted light microscope.

*Cytotoxicity of magnetic bone cement*

Cement discs (8 mm × 2 mm) were sterilized by $Co_{60}$ irradiation (25k Grey) and stored in a sterile environment. The CCK-8 assay (Dojindo, Japan) was performed to determine the viability and proliferation of L929 cells, co-cultured with their extracts on the PMMA cement and magnetic cement. The extracts of cements were prepared according to the ISO10993 standard[70, 71]. In brief, the aforementioned sterilized cement discs were incubated in MEM medium (cement sample to medium ratio: 0.2 g/mL) at 37 °C for 24 h. Meanwhile, cells were seeded into a 96-well plate at a density of $8 \times 10^3$ cells/mL and 100 μl growth medium per well. After being incubated for 24 h at 37 °C/5% $CO_2$, the medium solution was removed and replaced with 100 μl of the cement extracts in each well. The MEM without extracts was used as the control. At 1, 2, 3 days of culture, 100 μl of CCK reagent was added to each well respectively and the plate was further incubated at 37 °C for 2 h before the UV–vis absorbance measurement at 450 nm. Cell viability was expressed as a percentage of the control.

*In vivo temperature measurements in rabbits*

A bone cement rod containing 2 wt.% ZFO MNPs (aligned and random) was inserted into the medullary canal of the tibia of the rabbit, and exposed to an AC field of 380 kHz with varying amplitudes. The temperature of the medullary canal as a function of time of AC field heating was measured by a fiberoptic probe.

*Preparation of rabbit bone tumor model*

VX2 solid tumor mass was purchased from Tongpai Biological Technology Limited Company (Shanghai, China). Minced tumor tissues were injected into the muscles located in the right hind leg of New Zealand White Rabbits after anaesthetized by ketamine hydrochloride (10 mg/kg) and xylazine hydrochloride (5 mg/kg). After the tumor volume reached about 100 $mm^3$ in 20 days, the tumor masses were cut into small pieces to prepare for bone tumor model. VX2 solid tumors were transplanted in 20 New Zealand White Rabbits (20-24 weeks old) with body weight of 2.5-3.0 kg. All animal experiments were approved by the animal welfare ethics committee of Peking University Third Hospital. Operations of rabbits were performed with intramuscular injections of 10 mg/kg ketamine hydrochloride and 5 mg/kg xylazine hydrochloride. After incision of skin at knee joint, a 3 mm diameter hole was drilled on the tibial plateau parallel to the long axis of the tibia and then a magnetic bone cement rod of 3 mm in diameter and 35-40 mm in length was implanted into the bone marrow cavity. VX2 tumor fragments (3 mm$^3$) were implanted using puncture needle into the bone marrow cavity of the proximal tibia close to the tibial tubercle under the guidance of X-ray. Another short bone cement rod was used to seal the hole and the wound was sutured layer by layer.

*In vivo tumor magnetic hyperthermia treatment*

The tumor-bearing rabbits were divided into two groups (n=10 per group) randomly: magnetic hyperthermia treatment group and control group. For hyperthermia treatment group, the rabbits were exposed to an AC field with frequency of 380 kHz and amplitude of 7 kA/m for 30 min each day consecutively for four days.

**AUTHOR INFORMATION**

Corresponding Authors

*E-mail: shulihe@cnu.edu.cn

*E-mail: lvyang42@126.com

*E-mail: qingganggelin@126.com

*E-mail: haozeng@buffalo.edu

†These authors contributed equally to this work

**ORCID**

Shuli He: 0000-0002-0333-0553

Hao Zeng: 0000-0002-6692-6725

**ACKNOWLEDGMENTS**

This work was supported by National Natural Science Foundation of China (Grant No. 51771124, 51571146). Beijing Natural Science Foundation (Grant No. Z190011), Capacity Building for Sci- Tech Innovation - Fundamental Scientific Research Funds (Grant No. 20530290057).

# Bone tumor suppression in rabbits by hyperthermia below the clinical safety limit using aligned magnetic bone cement

Xiang Yu[1†], Shan Gao[2†], Di'an Wu[1†], Zhengrui Li[1], Yan Mi[1], Tianyu Yang[1], Fan Sun[3], Lichen Wang[1], Ruoshui Liu[1], Shuli He[1,#], Qinggang Ge[4,#], Yang Lv[2,#], Andy (Yuanguang) Xu[5], Hao Zeng[3,#]

[1] Department of Physics, Capital Normal University, Beijing 100048, China

[2] Department of Orthopedics, Peking University Third Hospital, Beijing 100191, China

[3] Department of Physics, University at Buffalo, SUNY, Buffalo, New York 14260, USA

[4] Department of Intensive Care Unit, Peking University Third Hospital, Beijing 100191, China

[5] Department of Radiation Oncology, Columbia University Medical Center, New York, NY 10032, USA

# 1. Synthesis of $Zn_{0.3}Fe_{2.7}O_4$ nanoparticles with sizes from 6-22 nm

ZFO MNPs with a diameter of 6 nm was synthesized by thermal decomposition in organic solution[1,2]. Briefly, Fe(III) acetylacetonate (2.7 mmol, 0.9536 g), Zn(II) acetylacetonate (0.3 mmol, 0.079 g), dodecanediol (10 mmol, 2.0234 g ), oleic acid (3 mmol, 0.96 mL), oleylamine (3 mmol, 1.02 mL), and 20 mL benzyl ether were mixed and magnetically stirred (500 rpm) under a flow of nitrogen (30 mL/min). The mixture was heated to 473 K at a rate of 5 K/min and kept at 473 K for 1 h. It was further heated to reflux (~ 573 K) at a rate of 10 K/min and kept at reflux for 2 h. The solution was cooled down to room temperature after removing the heat source. The MNPs were collected by adding ethanol and centrifugation (6500 rpm, 5 min), and were stored in hexane.

ZFO MNPs with a diameter of 9 nm were synthesized following a seed-mediated growth technique reported by Sun and Zeng[2]: briefly, Fe(III) acetylacetonate (0.9 mmol, 0.3179 g), Zn(II) acetylacetonate (0.1 mmol, 0.0263 g), dodecanediol (3.3 mmol, 0.6745 g), oleic acid (0.5 mmol, 0.16 mL), oleylamine (0.5 mmol, 0.17 mL), and 20 mL benzyl ether were mixed and magnetically stirred (500 rpm) under a flow of nitrogen (30 mL/min). The mixture was first heated to 393 K at a rate of 5 K/min and kept at 393 K for 30 min, and 100 mg of 6 nm MNP seeds were added. The mixture was then heated to 473 K at a rate of 5 K/min and kept at that temperature for 1 h. At a ramping rate of 10 K/min the solution was further heated and kept at 573 K for 2h. Following the purification procedures described in the synthesis of 6 nm particles, 9 nm ZFO MNPs were produced.

ZFO MNPs with a diameter of 12 nm were synthesized following the same seed-mediated growth technique using 9 nm MNPs as seeds, with identical seed to precursor ratio; ZFO MNPs with a diameter of 16 nm were synthesized following the same seed-mediated growth technique using 12 nm MNPs as seeds; ZFO MNPs with a diameter of 20 nm were synthesized following the seed-mediated growth technique using 16 nm MNPs as seeds; ZFO MNPs with a diameter of 22 nm were synthesized following the seed-mediated growth technique using 20 nm MNPs as seeds.

2. **Preparation of magnetic bone cement**

3.7 mg ZFO magnetic nanoparticles (MNPs) were dispersed in 0.75 ml MMA solution, and the dispersion were ultrasonicated in a water bath (90 W) for 10 min. As MNPs were coated by the surfactant of oleic acid and oleyamine, MNPs are highly dispersible in MMA. Then 1 g PMMA powder was added in MNP MMA dispersion. The mixture was mechanically stirred for 1 min. The mixture was injected into a cylindrical alumina crucible with a diameter of 10 mm, and a depth of 10 mm immediately. After about 40 min, the mixtures were solidified. For aligned magnetic bone cement, the solidification was proceeded under a DC magnetic field with a strength of 2 T.

## 3. TEM images and size distribution of ZFO MNPs with different sizes

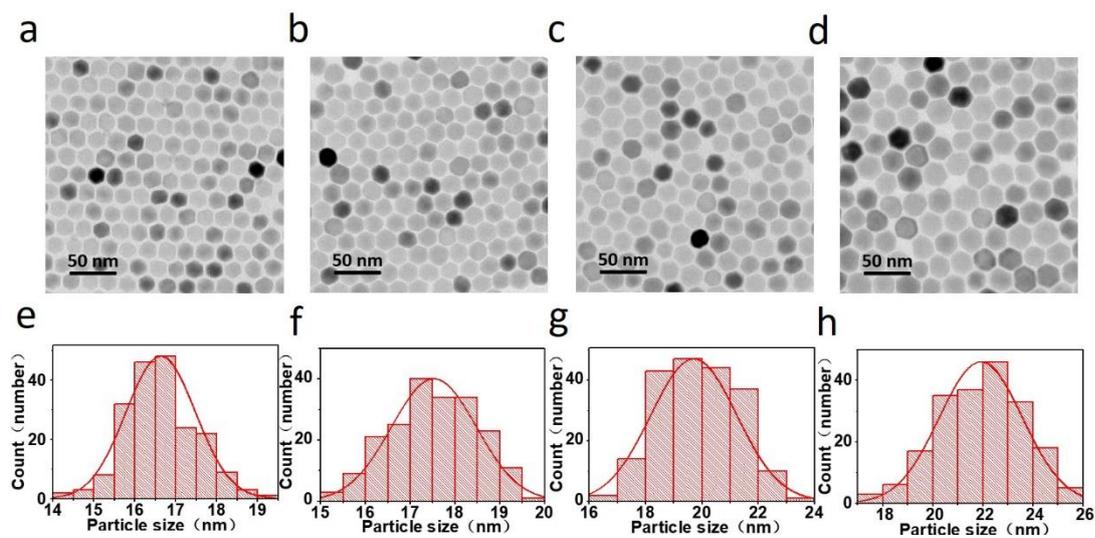

**Figure S1.** TEM images and size distribution of ZFO MNPs with different sizes (a) and (e)16 nm; (b) and (f) 18 nm; (c) and (g) 20 nm; and (d) and (h) 22 nm.

## 4. XRD patterns of the ZFO MNPs with different sizes

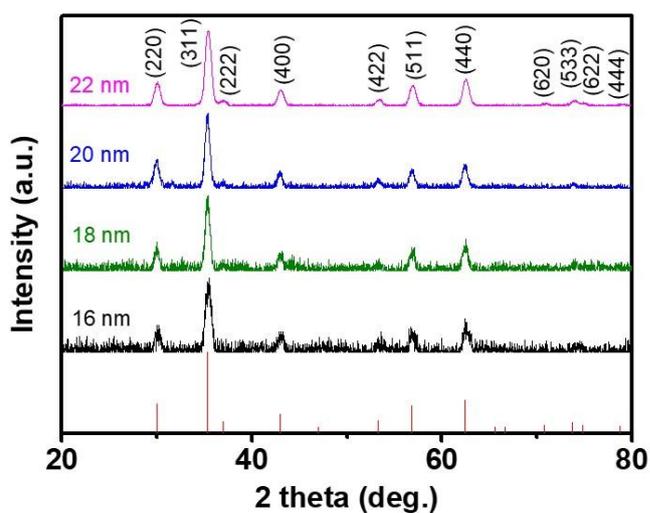

**Figure S2.** XRD patterns of the ZFO MNPs with different sizes.

5. **TEM images of magnetic bone cement with 22 nm ZFO nanoparticles**

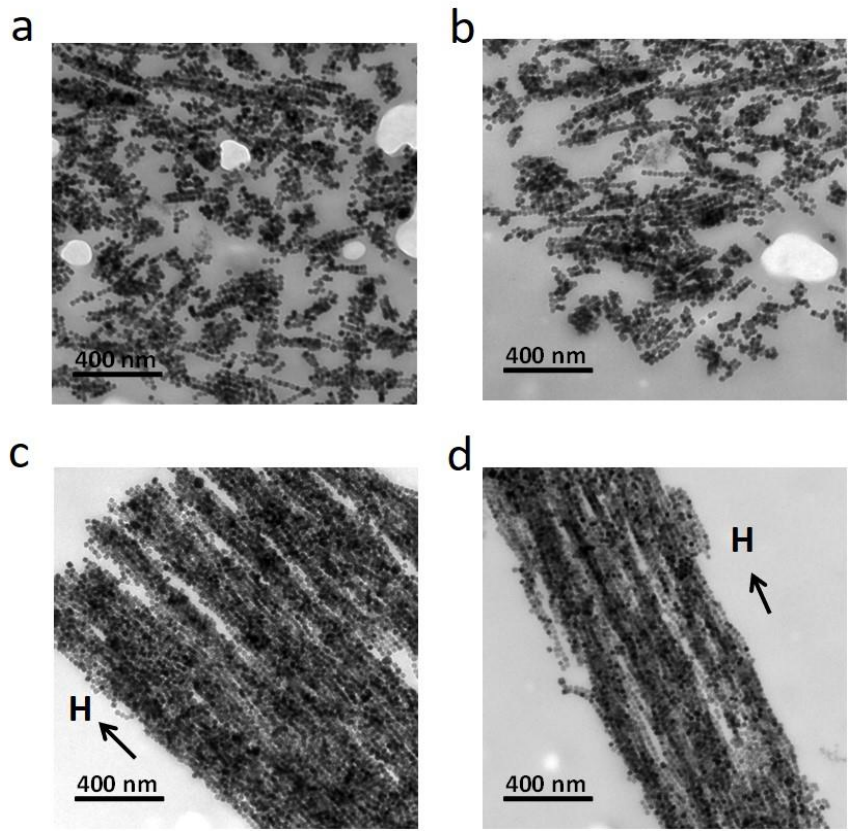

**Figure S3.** TEM images of (a) and (b) random; (c) and (d) aligned magnetic bone cement containing 0.2 wt.% 22 nm ZFO MNPs.

## 6. M-H curves of ZFO MNPs with different sizes

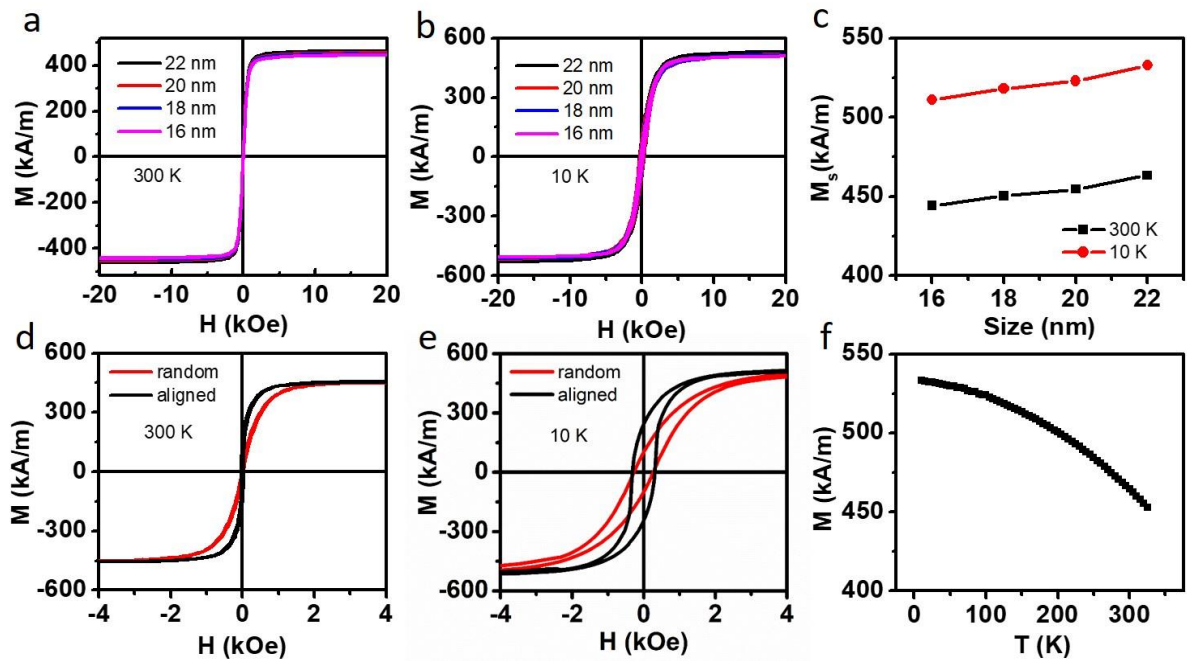

**Figure S4.** Hysteresis loops of ZFO MNPs with different size at (a) 300 K and (b) 10 K, (c) $M_S$ vs nanoparticle size. Hysteresis loops of aligned and random magnetic bone cement containing 0.2 wt% 22 nm ZFO MNPs measured at temperatures of (d) 300 K and (e) 10 K, (f) M-T curve of aligned magnetic bone cement containing 0.2 wt% 22 nm ZFO MNPs under a magnetic field of 10000 Oe, the vertical axis represents the magnetization of the ZFO MNPs.

7. **Heating curves of aligned and random magnetic bone cement containing 0.2 wt.% 22 nm ZFO nanoparticles**

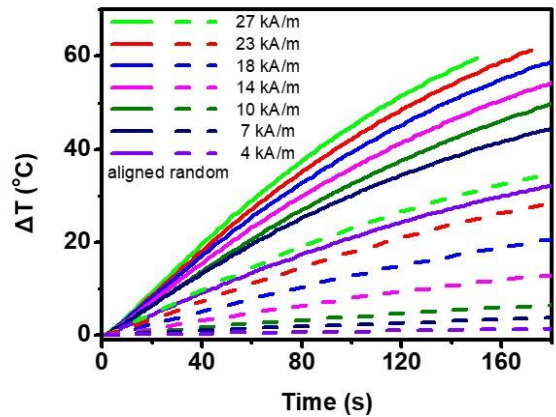

**Figure S5.** heating curves of aligned and random magnetic bone cement containing 0.2 wt.% 22 nm ZFO MNPs under the AC field of 430 kHz with different field amplitudes. Dashed lines are for random samples and solid lines are for aligned samples.

8. **Heating curves of aligned and random magnetic bone cement containing 0.2 wt.% ZFO MNPs with different sizes.**

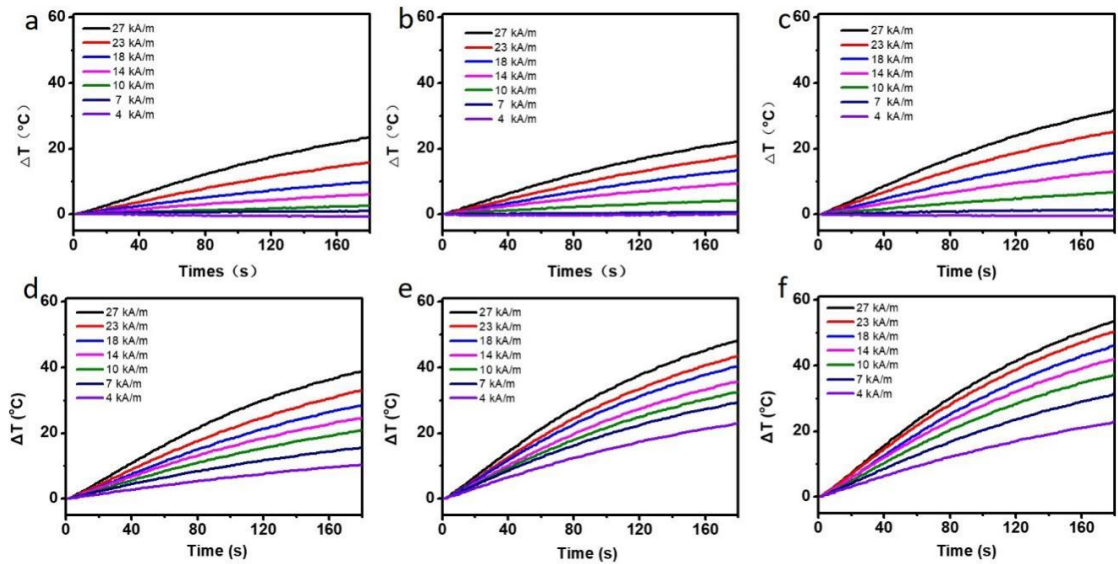

**Figure S6.** heating curves of (a) (b) (c) random and (d) (e) (f) aligned magnetic bone cement containing 0.2 wt.% of 16 nm, 18 nm and 20 nm $Zn_{0.3}Fe_{2.7}O_4$ MNPs under the AC field of 430 kHz with different field amplitudes.

9. **Optical images of the pig bone embedded with aligned and random magnetic bone cement**

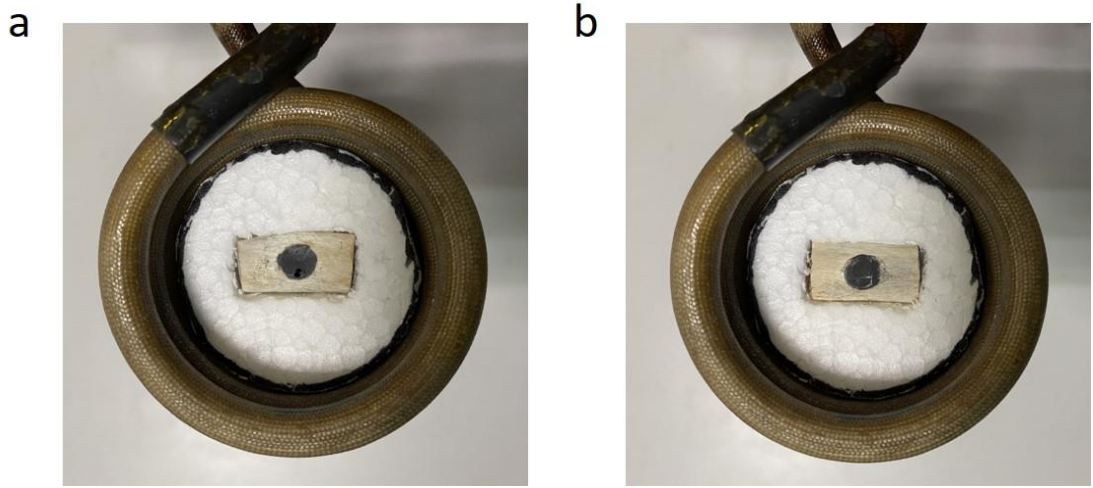

**Figure S7.** Optical images of the pig bone embedded with (a) aligned (b) random magnetic bone cement used for *in vitro* experiments.

## 10. Preparation of a tumor model in the rabbit bone marrow cavity

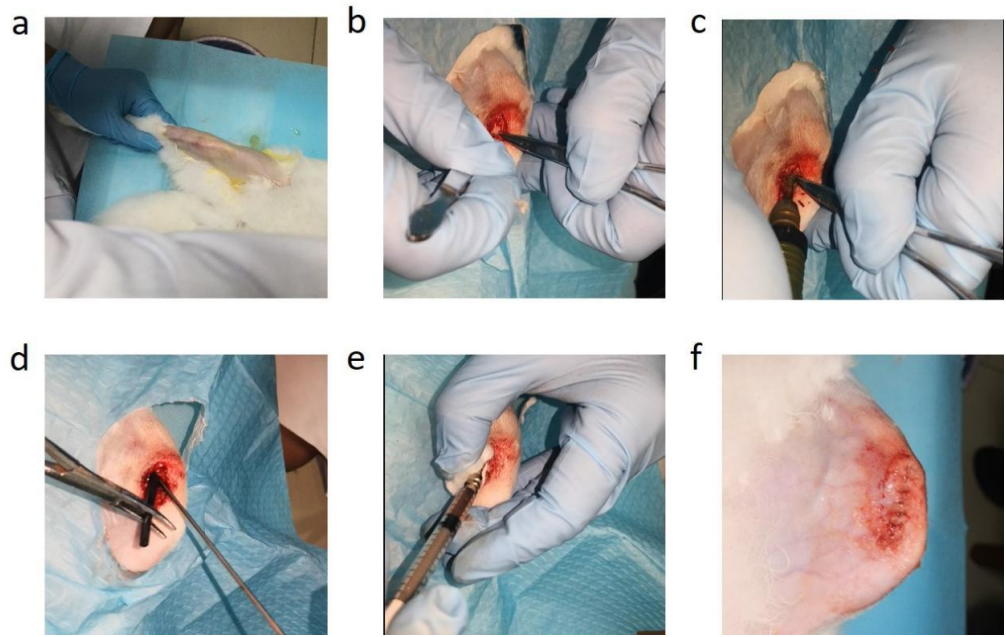

**Figure S8.** Preparation of a tumor model in the rabbit bone marrow cavity with a VX2 tumor mass. (a) Shaving and disinfecting the skin on rabbit knee joint. (b) Cut the skin, muscle and joint capsule cut layer by layer at the knee joint until exposing the tibial plateau. (c) 3 mm drill was used to prepare the bone defect at tibial plateau; (d) Implantation of bone cement rod; (e) 3 $mm^3$ fresh VX2 tumor tissue was injected into bone marrow cavity. (f) Suture the wound layer by layer.

## 11. MRI image of the rabbit tibia

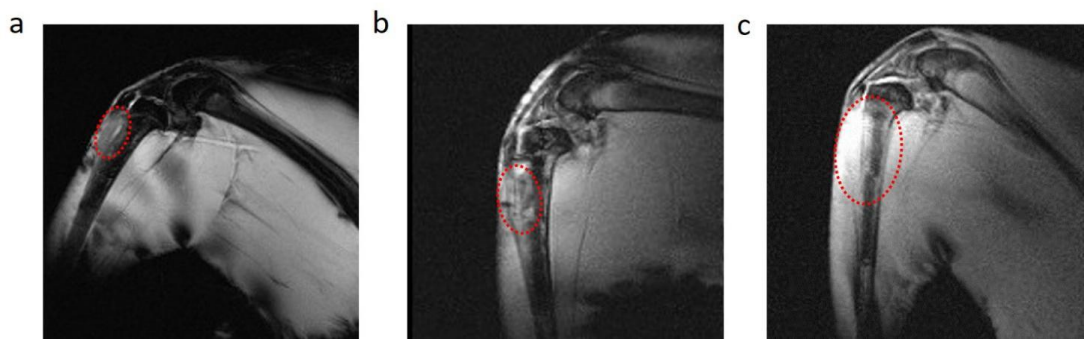

**Figure S9.** MRI images of the rabbit tibia after the tumor model implantation on the (a) first, (b) third and (c) fifth day.

## 12. Extraction of SLP values

For reliable extraction of SLP, all heating curves are fitted by the Box-Lucas formula $\Delta T = (S_m/k)(1 - e^{-k(t-t_0)})$ with $S_m$ and $k$ as the fitting parameters. $S_m$ is the initial slope of the heating curve, and $k$ is a constant describing the cooling rate. SLP is then calculated as $SLP = C_v S_m / \rho_i$, where $C_v$ is the specific heat capacity of the magnetic bone cement taken to be 1.579 J/(g·°C), and $\rho_i$ is the mass concentration of the metal in the magnetic bone cement (e.g. for $Fe_3O_4$, 1 $mg_{NPs}/cm^3$ = 0.724 $mg_{Fe}/cm^3$).

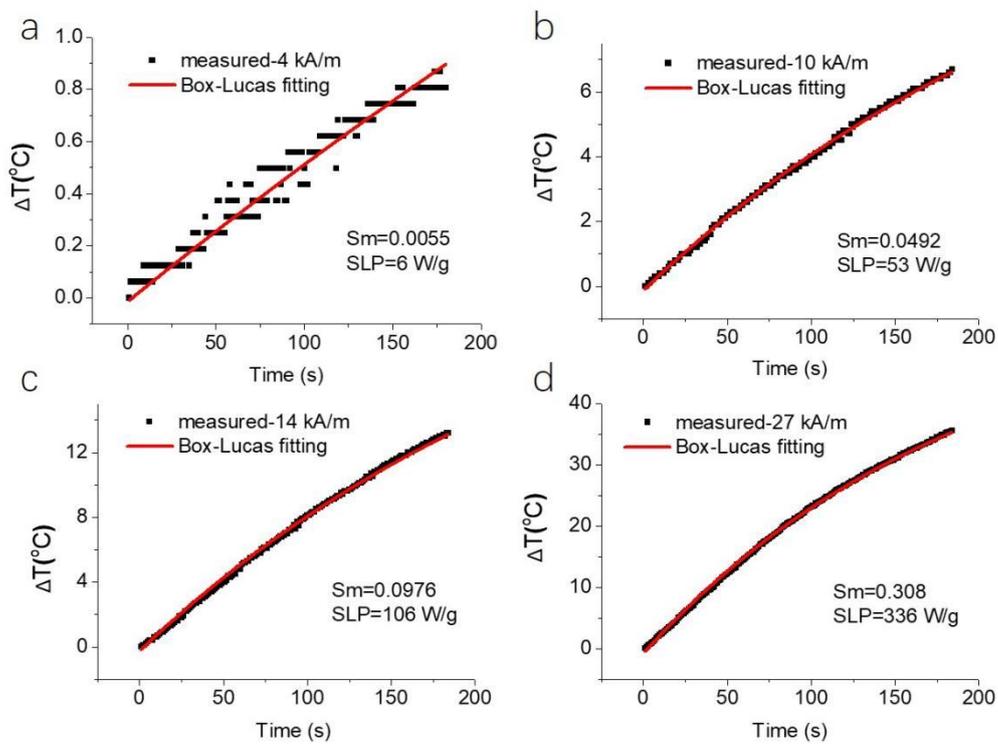

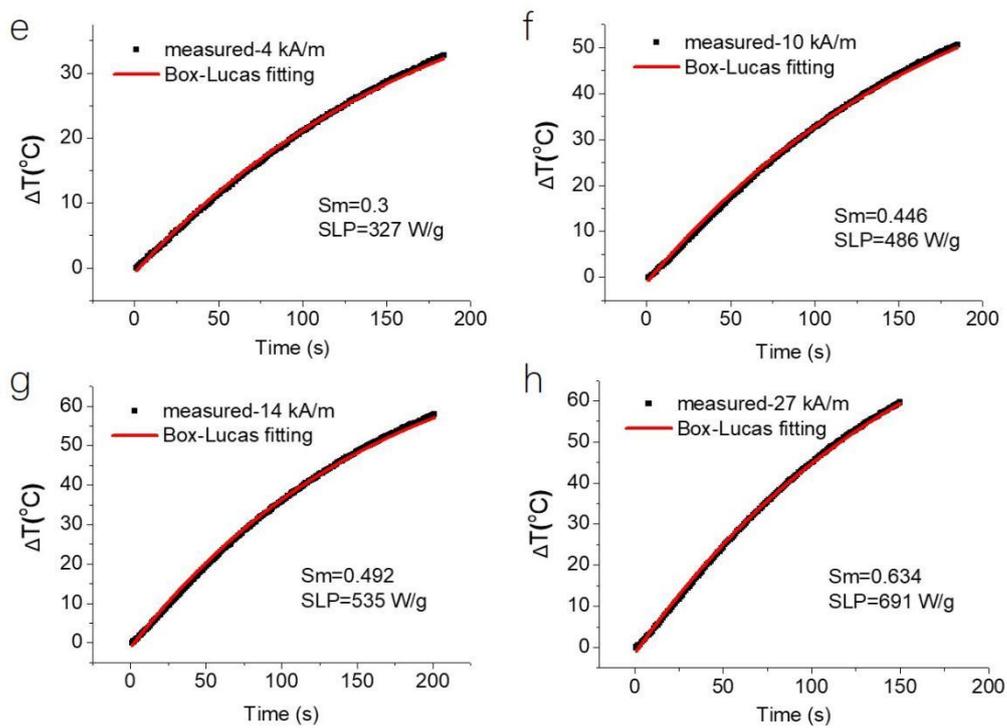

**Figure S10.** Examples of heating curve (black) and Box-Lucas fitting curve (red) of (a) (b) (c) (d) random and (e) (f) (g) (h) aligned magnetic bone cement containing 0.2 wt.% of 22 nm $Zn_{0.3}Fe_{2.7}O_4$ MNPs under the AC field of 430 kHz with different field amplitudes.

| random magnetic bone | | | | |
|---|---|---|---|---|
| H (kA/m) | 4 | 10 | 14 | 27 |
| dT/dt (K/s) | 0.0055 | 0.0492 | 0.0976 | 0.308 |

| aligned magnetic bone | | | | |
|---|---|---|---|---|
| H | 4 | 10 | 14 | 27 |
| dT/dt (K/s) | 0.3 | 0.446 | 0.492 | 0.634 |

## 13. Theoretical estimation of SLP

SLP can be estimated by A*f, where A is the area of the AC hysteresis loop and f is the frequency of the AC field. Since area of the AC hysteresis loops were not measured directly, it was obtained from low temperature DC hysteresis loops. According to S-W model for aligned particles (R. H. Victora, Phys. Rev. Lett. 1989, 63, 457),

$$H_C(T) = H_{C0}(T)\left[1 - \left(\frac{k_B T ln(f_0\tau)}{K_u(T)V}\right)^{1/2}\right] \quad (S1).$$

We assume if the calculated AC coercivity from Eq. (S1) at 300 K is the same as quasi-DC coercivity at a lower temperature T, they have the same minor hysteresis loop area (an assumption without strong justification, and hence such calculation is for trend prediction only).

$H_{c0}(0)$ and $K_u(0)$ were estimated from hysteresis loop measured at 10 K. $M_s(T)$ was measured experimentally; $H_{c0}(T)$ and $K_u(T)$ were extracted from (C. Zener, Phys. Rev. 1954, 96, 1335):

$$H_{C0}(T)/H_{C0}(0) \sim [M_S(T)/M_S(0)]^2 \quad (S2)$$

and

$$K_u(T)/K_u(0) \sim [M_S(T)/M_S(0)]^3 \quad (S3)$$

To the zeroth order, we first ignore the temperature dependence of $H_c$ and $K_u$, Eq.(S1) leads to

$$k_B T_1 ln(f_0\tau_1) = k_B T_2 ln(f_0\tau_2),$$

where $T_1 = 300\ K$, $\tau_1 = \frac{1}{f} = \frac{1}{430,000}$ $(s)$; $\tau_2 = 100\ s$. $T_2$ is then estimated to be ~ 93 K. Extracting $H_{c0}(T)$ and $K_u(T)$ at 93 K from Eq.s (2) and (3), plugging them into Eq. (1), we can obtain a new temperature $T_2$. The final $T_2$ value obtained by iteration method is ~ 100 K. We then use the minor hysteresis loop area (DC) measured at 100 K to represent the dynamic hysteresis loop area at 300 K and the loss power is then calculated as $SLP = A \cdot f$.

For example, at the temperature of 100 K, the minor loop area at 27 kA/m from numerical integration is $1.2 \times 10^{-3}$ J/g$_{metal}$. The estimated $SLP = A \cdot f = 1.2 \times 10^{-3} \times 430 \times 10^3 = 517$ W/g$_{metal}$.

## 14. SEM image and EDX mapped image of magnetic bone cement containing 0.2 wt.% ZFO nanoparticles

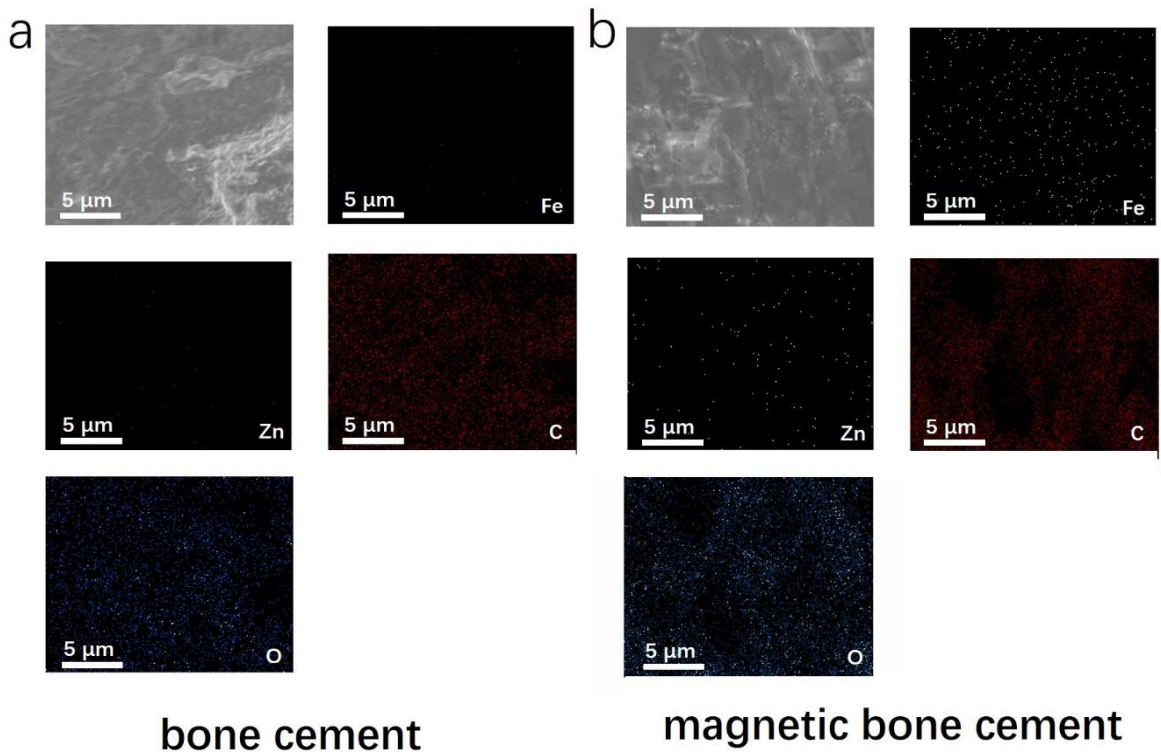

**Figure S11.** SEM images and EDX mapped images of (a) bone cement and (b) magnetic bone cement containing 0.2 wt.% ZFO nanoparticles.

## 15. SLP in recent studies (immobilized MNPs)

| Materials | Content of MNPs | ΔT (℃) | Time (s) | SLP (W/g) | AC field | Ref |
|---|---|---|---|---|---|---|
| Fe$_3$O$_4$-PMMA | 2 wt% | 20 | 200 | ~29* | 626 kHz, 5.72kA/m | 3 |
| | 4 wt% | 30 | 200 | ~60* | | |
| | 8 wt% | 80 | 200 | ~59* | | |
| Fe$_3$O$_4$-PMMA | 9 wt% | 60 | 180 | ~33* | 626 kHz, 5.72kA/m | 4 |
| Fe$_3$O$_4$/calcium phosphate bone cement | 40 wt% | 50 | 200 | 18-30 | 340 kHz, 11.3 kA/m | 5 |
| γ-Fe$_2$O$_3$@SiO$_2$-CaO heterostructures | 3.65wt% | 1.2 | 180 | 122 | 536.5kHz, 300 Gs | 6 |
| TiO$_2$-Fe$_3$O$_4$-PMMA | 25 wt% | 5 | 600 | - | 600 kHz, 40 Oe | 7 |
| | 15 wt% | 5 | 600 | - | 600 kHz, 100Oe | |
| Fe-akermanite ceramics | 2 wt% | 55 | 180 | - | 589 kHz, 0.768 kA/m | 8 |
| Fe$_3$O$_4$/GO bone cement | 10 wt% | 35 | 300 | - | 48 kHz, 250 Gs | 9 |
| Aligned Resovist-epoxy | 2.8 wt% | - | - | 10 | 100 kHz, 4 kA/m | 10 |
| | | | | 100 | 100 kHz, 16 kA/m | |
| Aligned Fe$_3$O$_4$ hydrogel | 0.2 wt% | 12 | 1000 | - | 400 kHz, 0.00256 kA/m | 11 |
| **Aligned Zn$_{0.3}$Fe$_{2.7}$O$_4$-PMMA** | **0.2 wt%** | **30** | **180** | **327** | **430 kHz, 4 kA/m** | **Our work** |

* These values are estimated from the heating curves provided in the paper.